
     
 \font\abstract=  cmr7 scaled\magstep1

 \magnification=\magstep1
\settabs 18 \columns
\hsize=16truecm

\def\b{\bigskip}
\def\bb{\bigskip\bigskip}

\def\no{\noindent}
\def\r{\rightline}
\def\ce{\centerline}
\def\ve{\vfill\eject}

\def\r{\rightline}

\def\harr#1#2{\smash{\mathop{\hbox to .25 in{\rightarrowfill}}
 \limits^{\scriptstyle#1}_{\scriptstyle#2}}}

\def\today{\ifcase\month\or January\or February\or March\or April\or
May\or June\or July\or
August\or September\or October\or November\or  December\fi
\space\number\day, \number\year }

\r \today
\bb\bb\bb




\def\sqr#1#2{{\vcenter{\vbox{\hrule height.#2pt
\hbox{\vrule width.#2pt height#2pt \kern#2pt
\vrule width.#2pt}
\hrule height.#2pt}}}}

 \def\1/2{{\scriptstyle{1\over 2}}}
 \def\a/2{{\scriptstyle{3\over 2}}}
 \def\5/2{{\scriptstyle{5\over 2}}}
 \def\7/2{{\scriptstyle{7\over 2}}}
 \def\3/4{{\scriptstyle{3\over 4}}}

\def\picture #1 by #2 (#3){
  \vbox to #2{
    \hrule width #1 height 0pt depth 0pt
    \vfill
    \special{picture #3} 
    }
  }

\def\scaledpicture #1 by #2 (#3 scaled #4){{
  \dimen0=#1 \dimen1=#2
  \divide\dimen0 by 1000 \multiply\dimen0 by #4
  \divide\dimen1 by 1000 \multiply\dimen1 by #4
  \picture \dimen0 by \dimen1 (#3 scaled #4)}
  }

%
%

\font\steptwo=cmb10 scaled\magstep2
\font\stepthree=cmb10 scaled\magstep4
\magnification=\magstep1

\ce{\stepthree Cosmology with an Action Principle}
\b
 \ce{Christian Fr\o nsdal}
\b
 \ce{\it Physics Department, University of California, Los Angeles CA
 90095-1547 USA}
\vskip1in

\def\sqr#1#2{{\vcenter{\vbox{\hrule height.#2pt
\hbox{\vrule width.#2pt height#2pt \kern#2pt
\vrule width.#2pt}
\hrule height.#2pt}}}}

\def \r{\rightarrow}

{\abstract  

\no{\it ABSTRACT.}   The Friedman universe is re-examined in a context
that is non-standard only in that the properties of  matter are postulated
in the form of an action principle.  
Applications to equilibrium configurations of ideal stars have already
been reported.
In this paper we apply the same theory to a fresh examination of the
Friedman universe. The results agree with standard theory in the case of
low densities. A suggestion is made to replace ``vacuum energy" by
``external force".}
\b\b
\no{\steptwo 1. Introduction}

To make this paper reasonably self contained, we present a brief summary
of the main idea and its merits, though it was done to the best of
our ability in [F].

Very briefly, this work is driven by a concern  about the standard
approach to relativistic dynamics that is presented in  the textbooks,
as a fusion of General Relativity and Nonrelativistic Thermodynamics. 
The basis for it all is Tolman's formula for the energy momentum tensor,
$$
T_{\mu\nu} = (\rho +p)U_\mu U_\nu - g_{\mu\nu} p,\eqno(1.1)
$$
used as the source of Einstein's field equations. It involves 2 scalar
fields, and a vector field $U$ that is dismissed   by an
argument that involves ``co-moving coordinates'',  and a normalization
condition that we believe had better been replaced by an equation of
motion. There is also a baryon density that seems curiously unrelated to
the rest.

Our approach takes General Relativity  with Einsteins equations for
granted, and tries to find model sources that have reasonable properties.
Classical thermodynamics is not posited as  input, instead
the interpretation is to be extracted from the
model itself. The fact that this interpretation can be understood in
thermodynamcal terms results from an examination of the structure of
the model, it is not postulated independently.
 
The four degrees of
freedom that are represented by the velocity field in the standard
treatment are not fully exploited in applications to  stellar dynamics
or cosmology. The space components are removed by specialization to
static phenomena, characterized by the absence of flow. 
 An alternative approach avoids introducing an independent
velocity field by focusing the attention on irrotational flows, as
first proposed by Jeans in 1902.   The possibility of describing
turbulence is thus discarded from the beginning; this is a strategy that
is standard in non-relativistic hydrodynamics whenever laminar flow is
expected to dominate, as in the theory of propagation of sound or heat. 
Thus in nonrelativistic hydrodynamics matter is described in terms of a
density  and a velocity potential. The equations
supposedly satisfied by these fields are Bernoulli's equation (derived
from Newton's equation) and the continuity equation. The theory admits a
formulation in terms of an action principle. The proposed relativistic
version of this theory involves a scalar field $\rho$  and a velocity
potential $\psi$,
 
$$
A_{\rm matter} =  \int d^4x \sqrt{-g}{\cal L} =\int d^4x \sqrt{-g} \Big(
{\rho\over 2}(g^{\mu\nu}\psi_{,\mu}\psi_{,\nu}-c^2)-  
V[\rho]\Big).
$$
The nonrelativistic theory is obtained by setting 
$$
\psi= c^2t   + \Phi
$$
and identifying $\Phi$ with the ordinary velocity potential. The
equations of motion are 
$$
{1\over 2}(g^{\mu\nu}\psi_{,\mu}\psi_{,\nu} - c^2) = {dV\over
d\rho}\eqno(1.2)
$$
and the equation of continuity,
$$
\partial_\mu(\sqrt{-g} \rho g^{\mu\nu}\psi_{,\nu}) = 0.\eqno(1.3)
$$
It is natural to interpret $V$ as an internal energy (including the
rest mass), this leads to an identification of the lagrangian density with
pressure 
$$
p = \rho{dV\over d\rho} - V = {\cal L} \,.
$$

Back  to the standard theory. After disposing of the  space
components of the velocity field it is the turn of the time component.
This embarrassing quantity is always eliminated, and always without
comment, by fiat: $g_{\mu\nu} U^\mu U^\nu = 1$! In the absence
of an action principle this does not lead to internal contradictions;
our discomfort arises from the fact that there is no justification and
every reason to be worried. The original context of this condition is the
motion of a particle along a geodesic. Here $U^\mu = dx^\mu/ds$ and 
$g_{\mu\nu} U^\mu U^\nu$ is a constant of the motion. Assigning the value
unity is the relativistic version of the identification $E = H$ in
classical mechanics. But the condition is not legitimate when more than
one particle is involved, unless each particle moves independently of the
others; that is, unless there is no interaction other than gravity. If
other forces are present it is absurd to impose this condition on each
particle separately. In fact, in relativistic particle theory one does
not know what to do. The Bethe-Salpeter equation is an unsatisfactory way
to deal with the problem, for strictly it makes sense only in the static
approximation. Relativistic mechanics of interacting particles does not
exist; the answer is field theory. Perhaps the best that can be said for
the condition $g_{\mu\nu} U^\mu U^\nu = 1$ is that, in the static case,
when it reduces to
$g_{00}(U^0)^2 = 1$, it is really irrelevant because this factor can be
absorbed into the density. True, but this merely transfers the problem
to the density;
which density is expected to satisfy an equation of state?  The
alternative approach replaces $U^\mu$ by
$g^{\mu\nu}\psi_{,\nu}$  and Eq.(1.2) constrains the invariant. In the
static case the derivative of this equation with respect to the radial
coordinate is exactly the same as the hydrostatic condition of the
standard approach. The only new element here is a new boundary condition
that turns out to be very beneficial in stellar dynamics. [F]

The other equation, Eq.(1.3), is a conservation law. In
the standard theory there is no conserved quantity. The density in
Eq.(1.1) is not conserved and its interpretation is controversial.  For
example, an equation of state relating $\rho$ to $p$, proposed by
Chandrasekhar  [C1], was taken over by other authors [OV] who  replaced
$\rho$ by the  energy density 
$\rho-p$.  The framework is sometimes expanded with the addition of a
conserved baryon density; it does not  have
natural place in the theory, but its mere existence was used to good
effect to justify the introduction of an equation of state [C2]. 

Nevertheless, the conservation law (1.3) does not impose additional
constraints on the theory, for when Eq.(1.2) holds it is equivalent to
the contracted Bianchi identity.

From now on, in this paper we are using units such that $ c = 1$.

\b
\ce{\bf Summary and conclusions}

In Section 2 we derive the equations that govern the metric and
matter fields in the Friedman universe. The metric is assumed to be such
that, in a suitable coordinate system, it takes the form of the
Robertson-Walker metric. As in the standard theory we seek solutions
without flow; that is,
$\psi$ depends only on the (Robertson-Walker)  time. The equations that
result from this, for the metric and for the fields $\rho, p$ and $\psi$,
would agree with the corresponding equations of the standard theory if
one would replace the equation
$$
{1\over 2}(\dot \psi^2 - 1) = {dV\over d\rho}\eqno(1.4) 
$$
by $\dot\psi = 1$ (and replace $\rho$ by $\rho + p$). Of course, the model
does not allow that, but it is a good approximation for very low pressure
and weak gravitational fields.

The time derivative of this equation (1.4) is a consequence of the 
contracted Bianchi identies and appears in the traditional approach as
well. Perhaps the most important property of the model is that Eq.(1.4)
is stronger than its time derivative; it fixes an additional constant. It
can be used to determine the value of $\dot\psi$ at the present time.
This will be done in Section 3 after selecting an equation of state.
 
In Section 3 we also present results of numerical integration of the
equations, forwards and backwards in time from present conditions, using
a polytropic equation of state.

The conclusion is that there is very little difference between the
predictions of the respective theories. The merit  of our proposal, seen
in the present limited context,  is, in our
opinion, aesthetic and improved logical coherence.   Most
remarkable is the insensitivity of the results with respect to wide
variation of the polytropic index. This  circumstance gives some
confidence that the results are not overly sensitive to the choice of the
equation of state.

In Section 4 we present a suggestion for overcoming the apparent
discrepancy between the observed density of matter of the universe
and the critical density favored by the measured value of the Hubble
constant and other recent data. 
\b
\ce {\bf More}
To include non-laminar flows, a generalization is needed, leading to
the difficult problems including turbulence. Accommodating rotating stars
and the Kerr metric is an easier task and it may serve as a first step
towards a general theory.
Quantization of the model, if it proves possible, may throw some light
on fluctuations. And thinking about this problem might possibly lead to
some inspiration about quantum gravity.
\b\b

\no{\steptwo 2. The  equations}

Einstein's equations,
$$
G_{\mu\nu} + \Lambda g_{\mu\nu} = 8\pi GT_{\mu\nu},
$$
 will be applied to the following situation.

It is assumed that there are coordinates $t,r,\theta,\phi$ in which the
line element takes the form
$$
ds^2 = dt^2 - R^2(t)\big( {dr^2\over 1-kr^2} + r^2 d\Omega^2\big).
$$ 
The components of the Einstein tensor depend only on $t$, so the same
must be true of the energy momentum tensor. The matter action is
$$
A_{\rm matter} = \int d^4x \sqrt{-g} \Big( {\rho\over
2}(g^{\mu\nu}\psi_{,\mu}\psi_{,\nu} -1) - V[\rho]\Big) = :\int
d^4x\sqrt{-g}\, {\cal L}.
$$
and the associated  energy momentum tensor  
$$
T_{\mu\nu} = \rho \psi_{,\mu}\psi_{,\nu} - g_{\mu\nu} \,{\cal
L}\,.\eqno(2.1)
$$
The fields $\rho,\psi$ are subject to the Euler-Lagrange equations
$$
\partial_\mu(\sqrt{-g} \rho g^{\mu\nu}\psi_{,\nu}) = 0,~~
{1\over 2}(g^{\mu\nu}\psi_{,\mu}\psi_{,\nu} - 1) = {dV\over d\rho}\,.
$$ 
We shall assume that both fields depend on $t$ only, then these equations
reduce to
$$
{d\over dt}(R^3\rho\dot \psi) = 0,~~ {1\over 2}(\dot\psi^2-1) = {dV\over
d\rho}.\eqno(2.2)
$$
The components of the Einstein tensors may be found in the textboks [W];
the nonzero components are
$$
G_{tt} =  (3/R^2)(\dot R^2 + k) , ~~~~G_{ij} = -R^{-2}(2R\ddot R
+ 
\dot R^2 +
 k)\tilde g_{ij} ,
$$
where $R^2\tilde g$ is the spatial part of the metric. Einstein's
equations boil down to
$$
(3/R^2)(\dot R^2 + k) -\Lambda = 8\pi G (\rho\dot\psi^2 - p)  , 
$$
and  one of the Bianchi identities (equation of   `energy conservation').
But the latter is implied by the wave equations and can be ignored.

The energy momentum density $\hat T_{\mu\nu} = \sqrt{-g}\,T_{\mu\nu}$
obtained from
 Eq.(2.1)(2.1) satisfies
$$
 (\hat T_\mu^\nu)_{,\nu} + \sqrt{-g}_{,\mu}{\cal L} + {1\over
2}\sqrt{-g}\,\rho \,g^{\alpha\beta}_{,\mu} \psi_{,\alpha}\psi_{,\beta} =
0,~~ \mu = t,r,\theta,\phi.
$$
In the present circumstances this reduces to one equation,  
$$
R^3\dot p ={d\over dt}(R^3 \rho \dot\psi^2).\eqno(2.4)
$$
If  we replace $\dot \psi$ by 1 in Eq.s (2.3) and (2.4),
and change the notation by translating $p\mapsto p,~\dot\psi^2\rho
\mapsto \rho + p$  we obtain the equations that define the Friedman
universe  according to the standard theory.

One can also derive the last result directly from the wave equations
(2.2). The first allows to define a constant $\kappa$,
$$
R^3\rho\dot\psi = \kappa,
$$
and the second gives upon derivation with respect to the time the result
$$
\dot p/\rho = \dot\psi\ddot\psi,\eqno(2.5)
$$
hence $R^3 \dot p   =  R^3\rho
\dot\psi\ddot\psi
=  \kappa \ddot\psi= (d/dt)(R^3\rho\dot\psi^2)$. 
  
To  summarize, the full set of equations is
$$
R^{-2}(\dot R^2 + k) = {8\pi G\over 3}(\rho\dot\psi^2 - p) +\Lambda/3,~~ 
R^3\rho\dot\psi = \kappa,~~ {1\over 2}(\dot\psi^2-1) = {dV\over
d\rho}
 \eqno(2.5-7)
$$ 
to be completed by a choice of the functional $V[\rho]$; that is, an
equation of state.  Note that (2.7) is stronger than (2.5)

In the traditional approach the equations are instead
$$
R^{-2}(\dot R^2 + k) = {8\pi G\over 3} \rho +\Lambda/3,~~~~ R^3\dot p
={d\over dt}\big(R^2(\rho + p)\big) = 0. 
 \eqno(2.8-9)
$$

Our result differs from the standard theory in several minor respects.
To begin with, Eq.(2.7) fixes $ \dot\psi$; that here plays the role of
$U^0$. It is not unity, but it does not differ significantly from
unity if the pressure is negligible. If we replace this equation by its
time derivative, then as was seen we end up with the standard set of
equations   by regarding the
energy density
$\rho\dot\psi^2-p$ as the fundamental variable and identifying it with
the field  $\rho$ of the standard theory.    The density used in the
equations of state are not the same, but that  is unimportant
if the pressure is negligible.

We pointed out that the Bianchi identity (2.4) can be derived from the
Euler-Lagrange equations (2.2) by differentiating the second one with
respect to to the time. The more important difference between our model
and the standard approach is that the equation
$$ 
{1\over 2}(\dot\psi^2-1) = {dV\over
d\rho}\eqno(2.10) 
$$
contains information that is lost upon differentiation. To understand
what is involved here let us consider the case of of a polytropic
equation of state, $V = a\rho^\gamma$, with $\gamma = 1 + 1/n$. Then this
equation reduces to
$$
{p\over \rho} = { \dot\psi^2-1\over 2n+2}.\eqno(2.11)
$$
Our model predicts the present value of $\dot\psi$ in terms of the
density and pressure. Unfortunately, the value exceeds unity by an
amount that is far too small to bridge the gap between observed density
and critical density. Some more discussion of this point may be found
in Section 4. 

 Finally, the appearance of $dV/d\rho$ in the equations has the effect
that, though we can eliminate the field $\dot\psi$, it is not   enough to
make a guess about the probable dependence of pressure and density on
$R(t)$. So we are in a somewhat more constrained situation in that we
shall need an equation of state. 

We shall explore the possibility that the universe is
a polytrope; that is, there is a constant 
$$
\gamma = 1 + {1\over n}
$$
such that
$$
V[\rho] = a\rho^\gamma,~~p = (a/n)\rho^\gamma.
$$
Then there is a function $f$ (the Emden function) such that
$$
V = af^{n+1}, ~~\rho = f^n,~~ p = (a/n)f\rho.
$$
The value of $f$ is closely related to the temperature. Eq.s (2.8-9)
become
 $$
R^{-2}(\dot R^2 + k) = {8\pi G\over 3}\rho\,(\dot\psi^2 - {af\over n})) +
\Lambda/3,~~  R^3\rho\,\dot\psi = \kappa,~~  \dot\psi^2-1  =
2\gamma af.
$$

Let $\rho_0, R_0,...$ denote the values of the variables at the present
time; we shall take $R_0 = 1$ and also $\dot R_0 = 1$, so that the unit of
time is the Hubble time $R_0/\dot R_0$. Then
$$
1 = {8\pi G\over 3}\rho_0(\dot\psi_0^2 - {af_0\over n} ) + \lambda  -k,
$$ 
where $\lambda$ is the reduced cosmological constant. Since the pressure
$p_0$ is neglible we have to a good approximation
$\dot\psi_0^2 - af_0/n  = 1$. Set
$ 
\rho = \mu \rho_{cr},
$ 
then 
$$
 k = \mu_0(\dot\psi_0^2 -  f_0/n) + \lambda  -1
$$ 
and our equations take the form that will be used for computations,
$$
\dot R^2 = \Big[\mu(\dot\psi^2 -af/n) + \lambda \Big]R^2 - k, ~~
R^3\mu\dot\psi = \mu_0\dot\psi_0,~~  \dot\psi^2-1 =2 \gamma a
f.\eqno(2.12-14) 
$$
Since it is difficult to estimate the value of the parameter $a$ we shall
set
$$
af(t) = bg(t),~~ g_0 = 1,~~ \mu/\mu_0 =  g^n . 
$$
and consider the constant $b $ as a free parameter.

We shall compare the results with the standard model, where the equations
are
$$
\dot R^2 = (\mu + \lambda )R^2 - k,~~ {\gamma\over n} R^3 \mu a\dot f =
{d\over dt}\big(R^3\mu(1+af/n)\big).
$$
Neglecting the effect of pressure gives complete agreement.

\b\b
\no{\steptwo 3. Numerical integration}

 We have already  set $ c = 1$. As unit of time we have taken Hubble time,
$R_0/\dot R_0$.
which is of the order of the current estimate of the age of the
universe, 18 billion years.   The parameters are
$ n,  \mu_0, \lambda,b   $. At time $t = 0$  the pressure  is
relatively unimportant, and the function
$\dot\psi$ is close to unity.     We begin our investigation with
the case $\mu_0 = 1, \lambda = 0,~b = 10^{-6}$, initial conditions
$af(0) = b, R(0) = 1$.

For small values of $a$, when the pressure is negligible at most times,
there is no difference between the standard theory and the model. Since
the pressure is negligible at present, and diminishes in the future,
there is no significant difference between the two theories in the
future. The function $R(t)$, renormalized is 1 at $ t = 0$ (present time)
and vanishes at time $-t_0$  (big bang).

The most interesting value of the polytropic index is $n = 3$,  which is
appropriate for a matter dominated universe and hence for most of the
past and all of the future. We start from the reference point
$$
(n, \mu_0,\lambda,b ) = (3,1,0,10^{-6})
$$
and vary one parameter at a time. At the refence point we have
$$
t_0 = -.666, ~~ q_0 = -\ddot R(t) = .5.
$$
We shall assume that experiment favors a greater age and a smaller
value of $q_0$.

\underbar{Increasing $n$} has
negligible effect, but smaller values of $n$ makes the universe younger.
Thus for $ n = .2$, $q_0$ remains at .5 and 
$$
t_0 = -.365,~~({\rm standard~theory)},~~~t_0 = -.596 ~~{(\rm dynamical~
model)}.
$$

\underbar{Increasing the pressure} has little effect unless it is
very high. Thus for $b = .1$,
$$
t_0 = -.611,~q_0= .56~~(\rm st.~ th.),~~~t_0 = -.575,~q_0= .67~~{\rm
(dyn. mod)}
$$
This is going in the wrong direction, but probably a reasonable mean
value for the parameter $b$ is much less than .1.

By current experimental evidence the value of $\mu_0$ is much less
than unity. We consider the alternative reference point
$$
  (n,\mu_0,\lambda,b) =
(3,.3,0,10^{-6}).
$$
 Then
$$
t_0 = -.809,~q_0= .15~~(\rm either~theory). 
  $$
After increasing the pressure as before, with $b = .1$,
$$
t_0 = -.741, q_0 = .2~~(\rm either~theory). 
$$
From the experimental point of view there is very little to distinguish
between the two theories. 
\b\b

\no{\steptwo 4. An alternative to vacuum energy}

Recent observations point to a spatially flat universe ($k = 0)$, 
while direct observation of the matter density gives only 30 percent of
critical.   The only popular way to resolve the  discrepancy is
to  attribute the missing 70\% to  ``vacuum energy"
represented by a positive  cosmological constant.  A 
theoretical prejudice against a positive cosmological constant (no
invariant vacuum and consequently no supersymmetry)  is a strong incentive
to look for alternatives.

The factor $\dot\psi^2$ that accompanies the density $\rho$ in Eq.(2.5)
seems to offer an opportunity, but it is
far too close to unity to be helpful. A generalization of the model may
be the answer.
\b

\ce{\bf Scale transformations in the traditional approach}

We have alluded to the fact that fixing the normalization of the
four-velocity field $U$ amounts to fixing the interpretation of the
density. Here we shall be more precise.

Let $q$ be a real, positive constant and define $\tilde U,\tilde
\rho,\tilde p$ by
$$
U = q\tilde U,~~ \rho = q^{-2}\tilde\rho,~~p = \tilde p.
$$
Tolmans formula goes over to $T_{\mu\nu} = \tilde \rho\tilde U_\mu\tilde
U_\nu - \tilde pg_{\mu\nu}$; it is scale invariant. The scale is,
nevertheless, meaningful. This for various obvious physical reasons and
also  since the transformation affects the equation of state. For
simplicity, suppose it has the form $ p = a\rho^\gamma$; then
$\tilde p =
\tilde a\tilde\rho^\gamma$, with $\tilde a = q^{-2\gamma}a$. If the value
of the constant $a$ is known, theoretically or experimentally, then it
implies a definition of density that fixes the scale. If we had no such
knowledge then the density would not be defined and setting either
$g^{\mu\nu} U_\mu U_\nu=1$ or $g^{\mu\nu} \tilde U_\mu \tilde U_\nu=1$
would be  equally justified.
\ve
\ce{\bf A cosmic force }

What would be the effect of adding a term proportional to the density to 
  our Lagrangian? We take
$$
{\cal L} = {\rho\over 2}\Big(g^{\mu\nu}\psi_\mu\psi_\nu - 1\Big) -
V -F\rho.
$$
In the nonrelativistic theory the addition of
the term proportional to the density is not observable, for it
corresponds to a sort of gauge transformation ([FW], page 301). If $\psi
= c^2t + \Phi$, then
$  g^{00}\dot\psi^2   -c^2 \approx  2c^2\Phi$. Here $\Phi$  is
the non-relativistic velocity potential; adding a constant to it does
not affect the velocity $\vec\bigtriangledown \Phi$.

The extra term can be eliminated by a scale transformation
$$
\psi = q\tilde\psi,~~ \rho =  \tilde\rho/q^2,~~ p = \tilde p.
$$
Our Lagrangian is not invariant; it takes the form  
$$
\tilde{\cal L} = 
{\tilde\rho\over 2}\Big(g^{\mu\nu}\tilde\psi_\mu\tilde\psi_\nu -
q^{-2}\Big) - \tilde V -(F/q^2)\tilde\rho.
$$
If we choose $q$ such that $F =  (q^2-1)/2$ then it becomes
$$\tilde{\cal L} = 
{\tilde\rho\over 2}\Big(g^{\mu\nu}\tilde\psi_\mu\tilde\psi_\nu -
1\Big) - \tilde V  .
$$
In the cosmological context the pressure is ill defined and usually
neglected in the case of a matter dominated universe. Consequently,
there is no independent determination of the external force or, what is
the same, the scale of the density. The wave equations now give, in the
case of a neglible pressure, $\dot{  \tilde  \psi}\approx   1 $. The
effective source for the gravitational field is $\tilde\rho 
=\rho/q^2$. If $ \rho$ is identified with the mass
density as measured locally, then a value $q^2 = 3$, or $F = 1$, will
appear to solve the problem of the  missing mass/energy and there will be
no need to invoke a non-vanishing cosmological constant.

Calling in a cosmic force is not a complete solution to  the problem,
since one would need to explain why it does not play a role  in the local
context.  The inverse is perhaps more plausible, a term proportional
to the density may manifest itself locally, leading to a low estimate of
the mass density.  The factor $q^2$ would then have the meaning of a
``biasing parameter" [SW].

\b

 \b
\b
\no{\steptwo References}
\settabs\+ ~~~~~~~&~~~&\cr
\+ [C1]& Chandrasekhar, S., Stellar Configurations with degenerate
cores,\cr 

\+ & Monthly Notices R.A.S, {\bf 95}, 226-260 (1935).\cr

\+[C2]&Chandrasekhar, S., The dynamical instability of gaseous masses
approaching the\cr 

\+&Schwarzschild limit in General Relativity,
Astrophys.J. {\bf 140}, 417-433 (1964).    
\cr

\+ [SW] &Strauss, M.A. and Willick, J.A., The density and peculiar
velocity fields of nearby\cr
\+& galaxies, Physics Reports {\bf 261}, 271-431
(1995).\cr

\+[F]&Fr\o nsdal,C.,  Ideal stars and General Relativity, UCLA
Preprint April 20, 2006.\cr

\+ [OV]&Oppenheimer, R.J. and Volkoff, G.M.,  On massive neutron cores,\cr
\+&Phys.Rev. {\bf 55},374-381 (1939).\cr

\end